\documentclass[epj]{svjour}
\usepackage{graphics}
\begin{document}
\title{Effect of screening of the electron-phonon interaction on the temperature of Bose-Einstein condensation of intersite bipolarons}
\author{B.Yavidov\inst{1,}\thanks{\emph{Permanent address: } Nukus State Pedagogical Institute named after A'jiniyaz,
230105 Nukus, Karakalpakstan, Uzbekistan} \and A.Kurmantayev\inst{2} \and D.Alimov\inst{2} \and B.Kurbanbekov\inst{2} \and Sh.Ramonkulov\inst{2} 
}                     
%
%
\institute{Institute of Nuclear Physics, 100214 Ulughbek, Tashkent,
Uzbekistan \and Ahmed Yasawi Kazakh-Turkish International
University, 161200 Turkestan, Kazakhstan}
\date{Received: date / Revised version: date}
%
\abstract{Here we consider an interacting electron-phonon system
within the framework of extended Holstein-Hubbard model at strong
enough electron-phonon interaction limit in which (bi)polarons are
the essential quasiparticles of the system. It is assumed that the
electron-phonon interaction is screened and its potential has
Yukawa-type analytical form. An effect of screening of the
electron-phonon interaction on the temperature of Bose-Einstein
condensation of the intersite bipolarons is studied for the first
time. It is revealed that the temperature of Bose-Einstein
condensation of intersite bipolarons is higher in the system with
the more screened electron-phonon interaction.
} 
\authorrunning{B.Ya.Yavidov, A.Kurmantayev\ldots}
\titlerunning{Effect of screening of the electron-phonon interaction\ldots}
\maketitle
\section{Introduction}
An electron-phonon interaction (EPI) in solids plays an important
role and in many cases determines their thermodynamic, electronic,
optical and other properties. The contribution coming from EPI to
the properties of solids is more pronounced in the case when the
coupling constant of EPI is much larger than those of the other
interactions. The latter may be an interaction between the same
charge carriers or an interaction between carrier and non-phonon
degree of freedom of the lattice. In solids, the {\it polaron}
concept has been widely used since the seminal paper of Landau
\cite{land}. The formation of a polaron quasiparticle in solids is
mainly owing to the presence of EPI. Though there are other
possibilities for the formation of a polaron via the other types of
interaction. Here we focus our attention only on EPI. Those polarons
which are formed by EPI are studied within the major frameworks: (i)
Fr\"{o}hlich model \cite{fro2}, (ii) molecular-crystal Holstein
model (HM) \cite{hol} and etc. In the first model a polaron forms
due to an interaction of an electron with the longitudinal optical
vibrations of a polar ionic crystalline. The crystal is assumed to
be a continuum. This means that one neglects the detailed structure
of the lattice. In the second case polaron formation is due to
coupling of a charge carrier to an intramolecular vibration of a
lattice. The Holstein model is commonly studied in a discrete
lattice. In the past both models have been extensively studied (see
for example reviewer papers \cite{shl-ston,amr,asa-devr} and books
\cite{firsov,bry,emin-pol-2012book}). A many polaron system in a
discrete lattice is often studied within the framework of
Holstein-Hubbard model
\cite{bonca-turgman-prb-64-094507-2001,magna-pucci-prb-55-14886-1997,macridin-prb-69-245111-2004},
extended Holstein-Hubbard model \cite{davenport-prb-86-035106-2012}
or Fr\"{o}hlich-Coulomb model
\cite{asa-kor-jpcm-14-5337-2002,hague-prl-98-037002-2007,hague-kor-prb-80-054301-2009}.
The models with extended interactions enable one to take into
account both the long-range feature of EPI and the correlation of
electrons at neighboring sites. At sufficiently strong EPI a many
polaron system is unstable with respect to the formation of a
bipolaron which is a bound state of two polarons. The conditions of
the formation of a bipolaron quasiparticle in solids were studied
extensively in the last decades for the purpose of pure academic
interest
\cite{bonca-turgman-prb-64-094507-2001,magna-pucci-prb-55-14886-1997,macridin-prb-69-245111-2004,davenport-prb-86-035106-2012,hague-kor-prb-80-054301-2009,takada-prb-26-1223_1982,verbist-prb-43-2712-1991,bassani-prb-43-5296-1991,smondyrev-prb-63-024302-2000,weibe-prb-62-r747-2000,wellein-prb-53-9666-1996,hohenadler-prb-71-184309-2005,golez-prb-85-144304-2012}
and for an interpretation of different phenomena, in particular in
the context of cuprates
\cite{asa-kor-jpcm-14-5337-2002,hague-prl-98-037002-2007,bassani-prb-48-12966-1993,iadonisi-pla-196-359-1995,s.dzhu-j.phys-c-254-311-1995,s.dzhu-prb-54-13121-1996,asa-prb-53-2863-1996,s.dzhu-j.ph.ch.sol-2012}
and alkali-doped-fullerides \cite{takada-hotta-ijmpb-12-3042_1998}.

A bipolaron is a boson. Therefore a bipolaron gas can, under certain
conditions, undergo Bose-Einstein condensation and thus would give
rise to bipolaronic superfluidity phenomenon (superconductivity).
Superconductivity of a charged ideal Bose-gas and superfluidity of
an ideal Bose-gas were first studied in
Ref.\cite{schafroth-pr-100-463-1955} and
Ref.\cite{blatt-pr-100-476-1955}, respectively. Bipolaronic
superconductivity is one of the mechanisms among others proposed for
the interpretation of high-$T_c$ phenomena in the cuprates. As the
problem of high-$T_c$ phenomena in the cuprates to date still
remains open an investigation of the properties of a bipolaron gas
may supply additional information about its relevance to the problem
of high-$T_c$ superconductivity of the cuprates.

The properties of a bipolaron gas are influenced by a number of
factors. For the intersite bipolaron \cite{on-site-bipolaron}  gas
these factors are: crystal structure, type of EPI, screening of EPI,
charge carriers' concentration, etc. Here we study only an effect of
screening of EPI on the temperature of Bose-Einstein condensation of
intersite bipolarons. On the one hand the issue is of considerable
academic interest for a broad community of (bi)polaron and
Bose-Einstein condensation physicists as such a task has not been
addressed so far. On the other hand the study is interesting from a
practical point of view as the obtained results can be used in a
wide range of phenomena \cite{griffin-BEC-book-1995}, in particular
in the cuprates \cite{asa-phys.scr-83-038301-2011} and metal-ammonia
solutions \cite{edwards-chem.phys.chem-7-2015-2006}.

In doing this we work with extended Holstein-Hubbard model and adopt
the analytical formula for the screened EPI introduced recently in
Refs.\cite{yav-pla-2010,yav-epjb-2010}.
\begin{figure}
\resizebox{0.5\textwidth}{!}{%
  \includegraphics{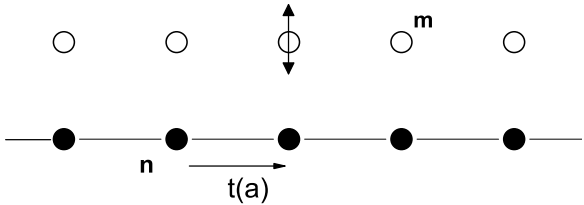}
}
\caption{In a one dimensional lattice of the extended Holstein model
an electron hops on the lower chain and interacts with vibrations of
the ions of an upper infinite chain via a density-displacement type
force $f_{{\bf m},\alpha}(\bf n)$. The distances between the chains
($b$) and between the ions of the same chain ($a$) are assumed equal
to 1.}
\label{fig:1}       
\end{figure}

\section{The model}
We consider an electron performing hopping motion on a lower chain
consisting of static sites, but interacting with all ions of an
upper chain via a long-range density-displacement type force, as
shown in Fig.1. The motion of an electron is always one-dimensional
and the upper chain's ions vibrate perpendicular to the chains. Such
a situation was studied in Ref. \cite{alekor} for polarons and in
Ref.\cite{bonca-turgman-prb-64-094507-2001} for bipolarons. Dynamics
of an electron (or hole) in such a model lattice {\it mimics} the
situation in the cuprates in which charge carriers belonging to
CuO$_2$ - plane and performing hopping motion along $a$- or $b$-
axes coupled with the {\it apical} ions. In the model lattice one
can show that small polarons have a small mass (as it was shown for
the first time by Alexandrov and Kornilovitch \cite{alekor}) and it
is also helpful for the explanation of dependence of $T_c$ (the
critical temperature of superconductivity) of cuprates on the
uniaxial strain (pressure) along $a$-, $b$- and $c$-
crystallographic axis
\cite{yav-physc-471-71-2011,yav-physb-407-2490-2012}. Here the model
lattice will be implemented for the study of the dependence of
Bose-Einstein condensation temperature of the ideal gas of intersite
bipolarons on the screening of EPI.

We write the Hamiltonian of the system of electrons and phonons
\cite{asa-kor-jpcm-14-5337-2002,alekor} as
\begin{equation}\label{Ham-full}
H=H_{e}+H_{ph}+H_{V}+H_{e-ph},
\end{equation}
where
\begin{equation}\label{H-e}
H_{e}=\sum_{{\bf n}\neq {\bf n'}}t({\bf n}-{\bf n'})c^{\dag}_{\bf
n}c_{\bf n'}
\end{equation}
describes the hopping of electrons between adjacent sites,
\begin{equation}\label{H-ph}
H_{ph}=\sum_{{\bf q},{\alpha}}\hbar\omega_{{\bf
q}\alpha}\left(d^{\dag}_{\bf q\alpha}d_{{\bf q}\alpha}+1/2\right),
\end{equation}
is the Hamiltonian of the phonon system,
\begin{equation}\label{H-Vc}
H_{V}=\sum_{{\bf n}\neq {\bf n'}}V_{C}({\bf n}-{\bf
n'})c^{\dag}_{\bf n}c_{\bf n}c^{\dag}_{\bf n'}c_{\bf n'},
\end{equation}
is the Hamiltonian of interacting particles at sites $\bf n$ and
$\bf n'$ via Coulomb forces, and
\begin{equation}\label{H-e-ph}
H_{e-ph}=\sum_{{\bf n}{\bf m}\alpha}f_{{\bf m}\alpha}({\bf
n})c^{\dag}_{\bf n}c_{\bf n}\xi_{{\bf m}\alpha}
\end{equation}
is the Hamiltonian of the electron-phonon interaction. Here $t(\bf
n-\bf n')$ is the transfer integral of an electron from site $\bf n$
to site $\bf n'$, $c^{\dag}_{\bf n}(c_{\bf n})$ is the creation
(annihilation) operator of an electron at site $\bf n$,
$d^{\dagger}_{\bf q\alpha}(d_{\bf q\alpha})$ is the creation
(annihilation) operator of a phonon with $\alpha$ ($\alpha=x,y,z$)
polarisation and wave vector $\bf q$, $\omega_{{\bf q}\alpha}$ is
the phonon's frequency, $V_{C}(\bf n-\bf n')$ is the Coulomb
potential energy of two electrons located at sites $\bf n$ and $\bf
n'$, $f_{\bf m\alpha}(\bf n)$ is the ''density-displacement'' type
coupling force of an electron at site ${\bf n}$ with the apical ion
at site ${\bf m}$ (Fig.1), and $\xi_{{\bf m}\alpha}$ is the normal
coordinate of ion's vibration on site ${\bf m}$ which is expressed
through phonon creation and destruction operators as
\begin{equation}\label{xi}
    \xi_{{\bf m}\alpha}=\sum_{\bf q}\left(\sqrt{\frac{\hbar}{2NM\omega_{{\bf
    q}\alpha}}}e^{i{\bf qm}}d^{\dagger}_{{\bf
    q}\alpha}+h.c.\right).
\end{equation}
Here $N$ is the number of sites and $M$ is the ion's mass. We work
with dispersionless phonons and take into account only the
$c$-polarised (perpendicular to the chains) vibrations of ions, as
charge carriers in the CuO$_2$ plane of the cuprates strongly
interact with $c-$ polarised vibrations of apical ions with
corresponding energy of the apical phonons 75 meV \cite{timusk}.
This apical phonon mode significantly contributes to polaron energy,
1.5 eV in La$_2$CuO$_4$ and 1.7 eV in YBa$_2$Cu$_3$O$_6$, and these
values of polaron binding energy suggest that EPI in the cuprates
are very strong \cite{gmzhao-prb-75-104511-2007}. Another evidence
of the fundamental role of EPI in the cuprates comes from the recent
pump-probe optical spectroscopy data
\cite{gadermaier-prl-105-257001-2010}. Early McQueeney and {\it et
al.} reported that coupling of charge carriers to the 75 meV mode
increases with increasing doping \cite{mcqueen-prl-82-628-1999}.
Theoretical treatment of doping dependence of EPI has also shown
that EPI remains significant at whole doping regimes and charge
carriers are in the non-adiabatic or near-adiabatic regimes
depending on doping level \cite{asa-brat-prl-105-226408-2010}.
Therefore our study will be restricted to two (non-adiabatic and
adiabatic) limits of strong EPI regime.

At strong coupling regime $\lambda=E_{p}/D\gg 1$ ($D-$ and $E_p-$
are the half-bandwidth and polaron shift energies, respectively) and
non-adiabatic limit $t/\hbar\omega<1$ one can use an analytical
method based on the extended (or nonlocal) Lang-Firsov
transformation and subsequently perturbation theory with respect to
the parameter $1/\lambda$ \cite{alekor}. It has been shown that
within the model (\ref{Ham-full}) an intersite bipolaron {\it{"can
tunnel from one cell to another via a direct single polaron
tunneling from one apex oxygen to its apex neighbor"}}
\cite{asa-mott-pol_and_bipolarons-book} and its mass has the same
order as polaron's mass \cite{asa-kor-jpcm-14-5337-2002}. For the
sake of simplicity we suppose that intersite bipolarons form an
 ideal gas of charged carriers and that the mass of the bipolaron is $m_{bp}=2m_p$ (this point does not lead to loose of generality). Then the temperature of
 Bose-Einstein condensation of an ideal gas of the intersite bipolarons is
 defined
 as (for details of derivation see \cite{lan-lif-theor.phys-v5-stat.phys})
\begin{equation}\label{tbec-formula}
 T_{BEC}=\frac{3.31\hbar^2n_{bp}^{2/3}}{2k_Bm_p}.
\end{equation}
Here $k_B$ is the Boltzmann constant and $n_{bp}$ is density of
intersite bipolarons. For the ideal gas of charge carriers one can
neglect interparticle Coulomb interaction, i.e. the term $H_V$
(Eq.(\ref{H-Vc})). In this case one can estimate polaron's mass
within the extended Holstein model (EHM) as \cite{alekor}:
\begin{equation}
    m_{non-ad-pol}=m^{\ast}e^{g^{2}},
\end{equation}
where $m^{\ast}$ is the electron's band mass and
\begin{equation}
    g^2_{non-ad}=\frac{1}{2M\hbar\omega^3}\sum_{{\bf m}}[f^2_{{\bf m}}({\bf
    0})-f_{{\bf m}}({\bf 0})f_{{\bf m}}({\bf 1})].
\end{equation}

For the determination of the mass of a small polaron at strong
coupling regime $\lambda\gg 1$ and adiabatic limit $t/\hbar\omega>1$
we adopt the corresponding analytical formulas of Ref.\cite{asa-yav}
for the mass of adiabatic small polaron at strong coupling regime.
Namely we estimate Extended Holstein adiabatic polaron's mass as
$m_{ad-pol}\simeq\hbar^2/2a^2\Delta E$, where $\Delta E=\Delta
\exp(-g^{2}_{ad})$,
\begin{equation}
\Delta=\frac{\hbar\widetilde{\omega}}{\pi}\sqrt{\frac{E_{p}}{2\hbar\omega}\kappa^{3/2}}\left(1-\sqrt{1-\left(\frac{E_{p}}{2\hbar\omega}\kappa^{3/2}\right)^{-1}}\right),
\end{equation}
\begin{equation}
g^{2}_{ad}=\frac{E_{p}}{2\hbar\omega}\kappa^{1/2}\sqrt{1-\left(\frac{E_{p}}{2\hbar\omega}\kappa^{1/2}\right)^{-1}},
\end{equation}
and
\begin{equation}\label{e-polaron}
    E_{p}=\frac{1}{2M\omega^2}\sum_{{\bf m}}f^2_{{\bf m}}({\bf 0}).
\end{equation}

Here $\widetilde{\omega}=\omega\sqrt{\kappa}$ is the renormalised
phonon frequency, $\kappa=(1-1/\lambda^2)$ and $\lambda=E_p/(2t)$.

As one can see from equation (\ref{tbec-formula}) at $n_{bp}=const$
the temperature of Bose-Einstein condensation of intersite
bipolarons is mainly determined by the form of EPI force $f_{\bf
m}$({\bf n}) which is in general screened. For the screened EPI
potential we adopt a more general form of screened interaction
potential which is Yukawa-type potential
\cite{yav-pla-2010,yav-epjb-2010}:
\begin{eqnarray}\label{Yukawa-potential}
  U_{{\bf m}}({\bf n}) &=& \frac{\kappa}{(|{\bf n}-{\bf m}|^2+b^2)^{1/2}}\times \\
\nonumber   &\times& \exp\left[-\frac{\sqrt{|{\bf n}-{\bf m}|^2+b^2}}{R}\right],
\end{eqnarray}
where $\kappa$ is some coefficient and $R$ is the screening radius
measured in units of $|\bf a|$. Such a potential was used in order
to explain the small value of charge carrier's mass and infrared
absorbtion in the cuprates \cite{yav-epjb-2010}. Here we use the
Yukawa-type EPI potential to study the influence of screening of EPI
on the temperature Bose-Einstein condensation of intersite
bipolarons.

From the potential Eq.(\ref{Yukawa-potential}) one obtains an
analytical expression for the screened force of EPI corresponding to
the interaction of an electron on site {\bf n} with the ion's
vibration on site {\bf m}:
\begin{eqnarray}\label{yukawa-force}
\nonumber  f_{{\bf m}}({\bf n}) &=& \frac{\kappa b}{(|{\bf n}-{\bf m}|^2+b^2)^{3/2}} \left(1+\frac{\sqrt{|{\bf n}-{\bf m}|^2+b^2}}{R}\right) \\
   &\times& \exp\left[-\frac{\sqrt{|{\bf n}-{\bf m}|^2+b^2}}{R}\right].
\end{eqnarray}

The formulas (\ref{tbec-formula})-(\ref{yukawa-force}) are main
analytical results according to which, in the next section, we
discuss the dependence of $T_{BEC}$ on the screening radius $R$.

\section{Results and discussion}

We have calculated the values of the temperature of Bose-Einstein
condensation $T_{BEC}$ of the intersite bipolarons on different
values of the screening radius $R$ for two regimes: non-adiabatic
($t/\hbar\omega\ll 1$) and adiabatic ($t/\hbar\omega\gg 1$). The
results are visualised in Fig.2 and Fig.3 for non-adiabatic and
adiabatic regimes, respectively. In non-adiabatic regime the
calculation is performed at $k^2/2M\hbar\omega^3=8.51$ in order to
get $T_{BEC}$ comparable with $T_c$ (transition temperature to
superconducting state) of the cuprates (Fig.2). The  same is done
for the adiabatic regime ($t/\hbar\omega\gg 1$) at
$k^2/2M\omega^2=0.188$ (Fig.3A). In the latter case the calculations
are performed with the $\hbar\omega$=75 meV and $t$=0.2 eV
\cite{bi-eklund-prl-70-2625-1993}, in order to ensure the
fulfillment of the condition of adiabaticy ($t/\hbar\omega\gg 1$).

\begin{figure}
\resizebox{0.5\textwidth}{!}{%
  \includegraphics{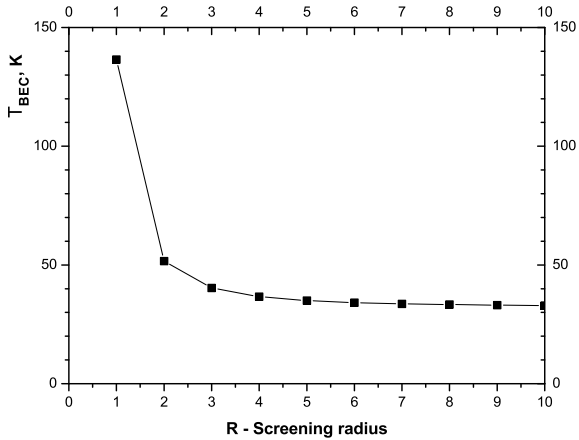}
}
\caption{The dependence of the temperature of Bose-Einstein
condensation of an ideal gas of the intersite bipolarons on the
screening radius $R$ of EPI at nonadiabatic regime and
$k^2/2M\hbar\omega^3=8.51$. The temperature is measured in Kelvin
and screening radius is given in units of lattice constant $a$.}
\label{fig:2}       
\end{figure}

\begin{figure}
\resizebox{0.5\textwidth}{!}{%
  \includegraphics{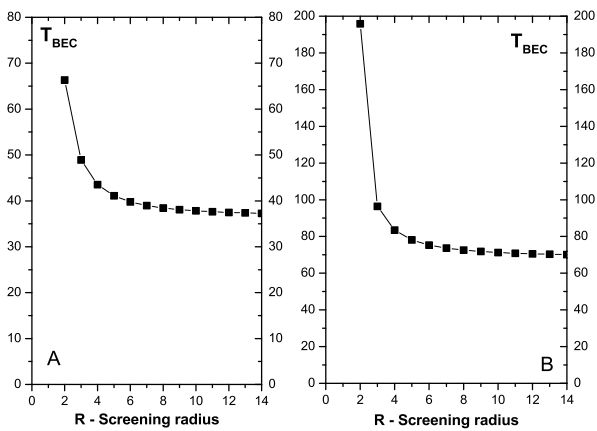}
}
\caption{The dependence of the temperature of Bose-Einstein
condensation of an ideal gas of the intersite bipolarons on the
screening radius $R$ of EPI in adiabatic limit and at
$k^2/2M\omega^2=0.188$ (Fig.\ref{fig:3}A), at $k^2/2M\omega^2=0.148$
(Fig.\ref{fig:3}B). The temperature is measured in Kelvin and the
screening radius is given in units of lattice constant $a$.}
\label{fig:3}       
\end{figure}

As one can see from the graphics the value of $T_{BEC}$ decreases
with the screening radius $R$, regardless of adiabaticy parameter
$t/\hbar\omega$. In both regimes the relative change of the value of
$T_{BEC}$ is more pronounced at the small values of screening radius
$R$. It seems that such a feature is hallmark of EHM as similar
behaviour has previously been observed for mass and optical
conductivity of polarons as well \cite{yav-epjb-2010}. In the
non-adiabatic regime a decrease of $R$ from $\infty$ to the value
$R=5$ increases the $T_{BEC}$ from $\simeq 32$ K up to $\simeq 35$
K, i.e. increased only by $\approx9$ \%. While at regimes of strong
screening of EPI, i.e. when screening radius is comparable to the
lattice constant, the increase of the value of $T_{BEC}$ is
considerably large. So the increase of the value of $T_{BEC}$ may
reach $\simeq 10$ K (increase by $\approx 28$ \%) or even $\simeq
85$ K (increase by $\approx 164$ \%) in the case of decreasing $R$
from $3$ to $2$ or from $2$ to $1$ respectively. A similar
dependence of $T_{BEC}$ on screening radius $R$ is observed in the
adiabatic limit too (Fig.3). An estimation of such a kind is
necessary when one studies charge carrier dynamics and their binding
(formation of a bound state of two carriers) at short distances
(i.e. within a few lattice units). It should be noted that the value
of $T_{BEC}$ in our model strongly depends on the model parameters
$k^2/2M\hbar\omega^3$ in non-adiabatic regime and $k^2/2M\omega^2$
in adiabatic regime. These parameters in turn depend on the
structure of cuprate through $k$, $M$ and $\omega$ and can take
values in a wide range. In order to illustrate this dependence, in
Fig.\ref{fig:3}B we plot $T_{BEC}$ versus $R$ in the adiabatic
regime for an another value of $k^2/2M\omega^2=0.148$. As it is seen
from the plot, $T_{BEC}$ can reach almost 195 K at values of $R$
comparable to the lattice constant.

It is also instructive to investigate the dependence of $T_{BEC}$ on
charge carrier's (in our case polaron's) concentration. In doing
this one distinguishes two regimes of screening of EPI. The first
regime is weak and moderate screening of EPI in which $R\gg1$ and
$T_{BEC}$ is nearly independent of $R$. The second regime is strong
screening of EPI, i.e. $R$ is comparable to the lattice constant,
$R\simeq a$. For the latter case, to simplify the consideration, one
can use a crude approximation to scale $T_{BEC}$ as $\sim R^{-1}$
(in both regimes of adiabaticy). It is appropriate to recall that
the screening radius itself depends on a number of parameters among
which is carriers' concentration. In the metallic regime of solids
the screening radius can be estimated within the Thomas-Fermi model
as $R_{TF}=(E_F/2\pi e^2 n_0)^{1/2}$, where $E_F$ - Fermi energy,
$n_0$- charge carrier's concentration. While in the semiconducting
regime one estimates the screening radius within the Debye model as
$R_{D}=(\varepsilon_0k_BT/4\pi e^2 n_0)^{1/2}$, where
$\varepsilon_0$ and $T$ are dielectric permittivity and absolute
temperature of a solid, respectively. As one can see in both models
$R$ scales as $\sim n_p^{-1/2}$. Here we replace $n_0$ with $n_p$
-polarons' concentration. Summarising, one writes (for a general
case)
\begin{equation}\label{Tbec-nbp-np}
    T_{BEC}\sim n_{bp}^{2/3}n_p^{1/2}.
\end{equation}

\begin{figure}
\resizebox{0.5\textwidth}{!}{%
  \includegraphics{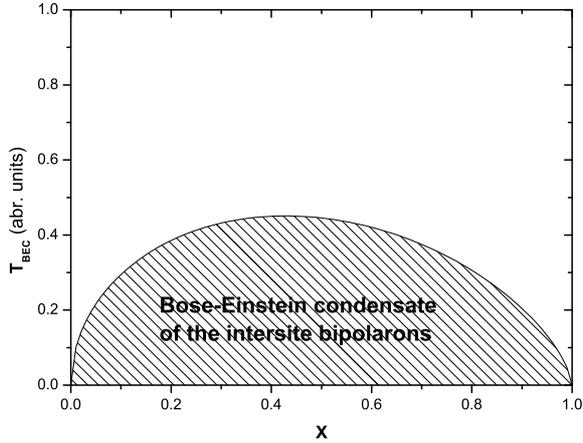}
}
\caption{Phase diagram of the ideal gas of intersite bipolarons at
strong EPI screening regime on ($x$,$T_{BEC}$) plane.}
\label{fig:4}       
\end{figure}

Taking into account that the total number of charge carriers in the
system is conserved $n_p+n_{bp}=N=const$ one rewrites the
Eq.(\ref{Tbec-nbp-np}) as
\begin{equation}\label{Tbec-x}
    T_{BEC}\sim (1-x)^{2/3}x^{1/2},
\end{equation}
where $x=n_p/N$. In Fig.\ref{fig:4} the latter relation between
$T_{BEC}$ and $x$ is presented graphically.  The relation
(\ref{Tbec-x}) separates ($x$,$T_{BEC}$) plane into two regions. An
area below of the line (\ref{Tbec-x}) represents Bose-Einstein
condensate of an ideal gas of intersite bipolarons. In the area
above the line Bose-Einstein condensation of intersite bipolarons is
impossible. Thus Fig.\ref{fig:4} is a phase diagram of an ideal gas
of intersite bipolarons. The form of the area of Bose-Einstein
condensate of intersite bipolarons as seen from Fig.\ref{fig:4} is
slightly asymmetric and shifted to the low values of $x$. It is
trivial that there is no Bose-Einstein condensate at $x=0$ (no
polarons and consequently no bipolarons) and $x=1$ (all charge
carriers are polarons). At some point determined by the condition
$\partial T_{BEC}/\partial x=0$ the temperature of Bose-Einstein
condensation of intersite bipolarons reaches a maximum. The point is
$x=3/7$. It is worthwhile to notice that this phase diagram is
obtained under crude approximations and represents a particular case
of screening regime of EPI and its effect on $T_{BEC}$ of intersite
bipolarons. However, the model in which polarons and bipolarons can
coexist, and EPI plays an important role, seems to be a plausible
model for explaining qualitatively (i) high values of $T_c$
(=$T_{BEC}$) and (ii) "bell-shaped" form of the dependence of $T_c$
(=$T_{BEC}$) versus charge carrier's concentration, which in our
case is polarons, $n_p$. Previously, a system in which large
polarons and large bipolarons coexist, was studied in the context of
high-T$_c$ superconductivity and typical behaviour of $T_c$ versus
$x$ found in the cuprates was reproduced in Ref.
\cite{iadonisi-pla-196-359-1995}. Such a system of polaron-bipolaron
mixture, and Bose-Einstein condensation of bipolarons in such a
system deserves a more detailed consideration as polarons in this
mixture stabilise the isolated nonstable large bipolaronic states
\cite{smondyrev-prb-63-024302-2000}. This is the very case that is
considered in this paper. Though we study here an ideal gas of
intersite bipolarons, our calculation of $T_{BEC}$ indirectly takes
into account screening of the Coulomb interaction, which is
responsible for the doping dependence of the bipolaron binding
energy and its effective mass
\cite{asa-brat-prl-105-226408-2010,asa-brat-epl-95-27004-2011}. In
this work such an account is done by expressing $T_{BEC}$ through
the screening radius $R$ via the mass of a bipolaron $m_{bp}=2m_p$.

As a rule, when screening is increased, the effects of EPI become
weaker, and consequently the probability of the formation of
bipolarons decreases as well. Therefore one must bear in mind that
our study is valid only to the case when bipolaron formation is
possible. For a comprehensive study of the problem with all aspects
one should take into account nonideality of Bose-gas or other
factors in the system. Study of Bose-Einstein condensation of
nonideal Bose-gas may be useful in Fermi-Bose-liquid scenarios of
superconductivity as well
\cite{s.dzhu-pramana.j.phys,s.dzhu-j.phys-c-235-2269-1994,s.dzhu-ijmp-b-12-2151-1998}.
Consideration of the more general cases is in progress.

\section{Conslusion}

In conclusion, we have studied the effect of screening of EPI on the
temperature Bose-Einstein condensation of an ideal gas of intersite
bipolarons. It is shown that the effect is more pronounced at the
values of screening radius comparable with the lattice constant. The
screening of EPI gives rise the increase of the temperature of
Bose-Einstein condensation of the intersite bipolarons. The
here-revealed feature of an ideal gas of intersite bipolarons i.e.
dependence of its Bose-Einstein condensation temperature on
screening radius should be taken into account when applying this
scenario of superconductivity to real systems.

\begin{acknowledgement}
The author (B.Ya.Ya.) acknowledges financial support from the
Fundamental Research Programme of Uzbek Academy of Sciences (grant
no. ${\Phi}$2-${\Phi}$A-${\Phi}$120) and from the Fund for Basic
Research of Uzbek Academy of Sciences (grant no. $\Phi$.2-12).
\end{acknowledgement}

%

\begin{thebibliography}{99}
\bibitem{on-site-bipolaron} {\small On-site bipolaron (bound state of two polarons on the same site of a lattice) is ruled out from consideration
 due to its huge mass and therefore its irrevalance to the cuprates. Here we consider only intersite bipolaron which is bound state of two polarons sitting on the neighboring sites of a lattice.}
\end{thebibliography}
%

\end{document}